# Use of BlockChain in the Internet of Things: A Survey

Mohammad Qatawneh

The University of Jordan,Jordan
Department of Computer Science,
King Abdullah II School for Information Technology
mohd.qat@ju.edu.jo

**Abstract.** The Internet of Things (IoT) devices is being implemented in many fields like military, health care, smart environments, etc. where such devices or things may gather, process, and upload a huge amount of data to the internet, which can lead to many challenges in terms of keeping IoT secure, protecting the IoT from disclosure to unauthorized parties, ensuring information integrity and so on. Such challenges can be addressed by BlockChain (BC) technology. This paper provides a comprehensive and high-level technical overview of BC technology by creating a common language through business and technology about BC to adapt it to the specific needs of IoT in order to develop BC-based IoT (BC-IoT) applications. Then an overview of a BC technology, its architectures, classification, challenges, etc. is presented. Finally, some recommendations are provided to researchers that will have to be tackled before deploying the next generation of IoT applications based on BC technology.

## 1 Introduction

The Internet of Things (IoT) can be defined as an information system made up of things or subjects, networks, data, and services. Such things may be wireless sensors, traditional computers, cameras, home appliances, tablets, smart phones, vehicles, humans, etc. that are connected over a network that can be wired or wireless. These things may gather, process, and upload a huge amount of data to the internet and used to initiate service. It can be also defined based on the type of technology used as a combination of several technologies: communication, information and embedded technologies that make up an IoT ecosystem.

Currently, most Internet of Things solutions based on centralized architecture. Therefore, the exchange of information and data authentication in the IoT system is only done via the central server, connecting to cloud servers via the Internet, thus leading to security and privacy concerns especially with the growing number of IoT devices that will be used by 2025 [1][2][3][15]. The third trusted party (central server) which guarantees safe and secure delivery between two parties is questionable in case of modifying the original data, delays in delivery, fraud, or vulnerable to system failures and hacking [4][14][34][35][36], because with just a single point of control responsible for security many questions arise such as what if this third party becomes untrusted in terms of that the data may be updated or the system can be a victim of considerable hacked by cybercriminals [5][6][7][12]13]. Other questions can arise such as can we use the p2p communication paradigm to tackle the delay in communication as well as the issue of authentication of transactions.

Consequently, new architectures will have to be proposed to overcome such concerns. Among such suggestions, decentralized models were suggested to create large P2P Wireless Sensor Networks [8], but some pieces were missing in relation to privacy and security until the arrival of BC technology [9][10][11][37]. The BlockChain (BC) technology, which was invented by Satoshi Nakamoto in 2008 to underpin bitcoin can be a panacea for many challenging problems across many businesses and industries. BlockChain technology promises to address many issues such as security, privacy etc. due to several attractive features that it has.

The goal of this paper is to provide a comprehensive and high-level technical overview of BC technology to help students, readers, and interested understand what does BC mean, how does BC work, how BC relates to Bitcoin, what kinds of problems it solves, and what are the differences between BB and Bitcoin, because there is a common misconception among many people that BC and bitcoin are one and the same. In addition to that a comprehensive review on how to adapt BC to the specific needs of IoT in order to develop BlockChain-based IoT applications is presented. The rest of the paper is organized as follows. Section 2 presents the cross-border money transfer problem and the definition of BC. Section 3 describes the basics of BC technology: its architecture, challenges, applications, classifications, and how does it work. Finally, section 4 concludes the paper.

## 2. Conventional Cross-Border Money Transfer Problem.

Before exploring the details of BC, it is important first to explain why BC was invented and how it can underpin Bitcoin by discussing how does the Cross-Border Money Transfer Process work before and after emerging the BC technology. Let's start how the money can be moved from one account to another by traditional cross-border money transfer process. The conventional cross-border money transfer process can be made via SWIFT system as shown in Fig. 1. The workflow of transferring money from sender to receiver is given below:
- Bank A sends the payment request to its correspondent bank B via settlement instructions in sending country.
- Bank B does the settlements and sends a message to its correspondent bank C.
- Bank C transmits the money to the final beneficiary in bank D.

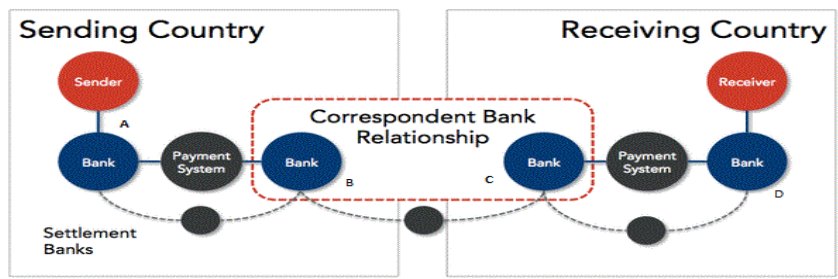

**Fig. 1**: The conventional cross-border money transfer process [16]

As shown in the workflow of transferring money and in Fig.1, the conventional money transfer way has the following three shortcomings:
- The process of transferring money is done via a third trusted party (set of banks A, B, C, and D).
- The third trusted party (banks) charges fees for processing of each transaction and this adds to the cost and thus increasing the cost.
- The process of transferring money takes time about 2 or 3 days to send money from the sender to the receiver.

There should be a novel technology to overcome the above shortcomings of the conventional cross-border money transfer process, BlockChain technology comes into play.

### 2.1 Cross-Border Money Transfer Using BlockChain.

The BC technology was invented in 2008 to underpin bitcoin, the first digital cryptocurrency to address the shortcomings of the conventional money transfer way as discussed in previous section by enabling moving digital currency (bitcoin) from one account to another account by eliminating the need of third trusted party, which leads to reduce the time and cost (fees) needed to move money via traditional way. In addition to that BC provides better security than a conventional approach, which means less of the fraud that could happen when using the conventional approach. The better security is achieved because the data or transactions recorded in the blocks can't be modified due to the fact that BC uses private and public keys for encryption, decryption, hashing and digital signature as will be discussed in the next section. The differences between conventional and BC money transfer approaches are shown in Table 1.

**Table 1.** The differences between conventional and BC money transfer approaches.

|  | Conventional Way | BlockChain Technology |
|---|---|---|
| Time | The time needed to transfer can take up to two to four days to complete. | Immediately. |
| Third Party | There is a need for third party. | No need for any third party. |
| Cost | The intermediary banks charge fees. | Cheaper than the conventional way. |

This section presents how BC technology facilitate the process of cross-border money transfer by eliminating the third party, faster time, and cheaper than the conventional way through explaining the following three principles of BC technology: Open Ledger, Distributed, and Mining.

1.  **Open Ledger**: Open Ledger is an important concept of BC technology that needs to be explained in order to understand how does BC work. Consider a system consisting of four nodes A, B, C, and D and node A has ten dollars (A=$10 - Transaction 0). The concept of open ledger can be explained as follows:

    - Assume that node A wants to send five dollars to node B (A →B $5- Transaction 1), then this transaction will be added and linked into the transactions' chain as shown in Fig. 2.
    - Suppose that node B wants to send three dollars to node C (B → C $3- Transaction 2), this transaction is also will be added and linked into the chain of the ledger as shown in Fig. 2.
    - Finally, consider that node C wants to send one dollar to node D (C→ D $1- Transaction 3), this transaction is also will be added and linked into the existed transactions (0,1, and 2) as shown in the Fig. 2. Now, all transactions (0,1,2,3) are linked into the chain, which will be broadcasted to all other nodes in system as shown in Fig. 3.

2.  **Distributed Ledger**: The transactions (0,1,2,3) are broadcasted to all nodes in the system as shown in Fig.3. Thus the nodes (A, B, C, and D) have the same copy of the digital ledger. In this context, it is important to mention that, there is no need for a central node to hold all the ledger as shown in Fig. 3 and 4. Based on the distributed ledger every node in the system can decide that the transaction is valid or not valid. For example, if the node A tries to send fifteen dollars to node D, then the nodes decide that this transaction is not valid because node A does not have enough money to send, and this transaction will not be added to the ledger as shown in Fig. 5.

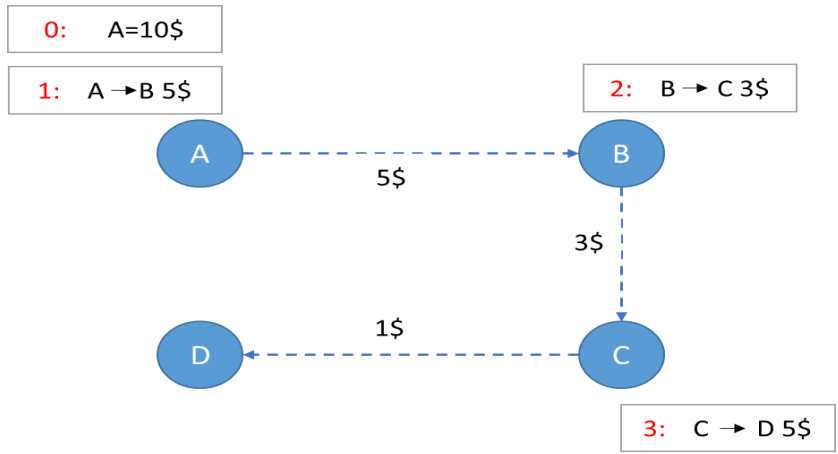

**Fig. 2**: The concept of open ledger

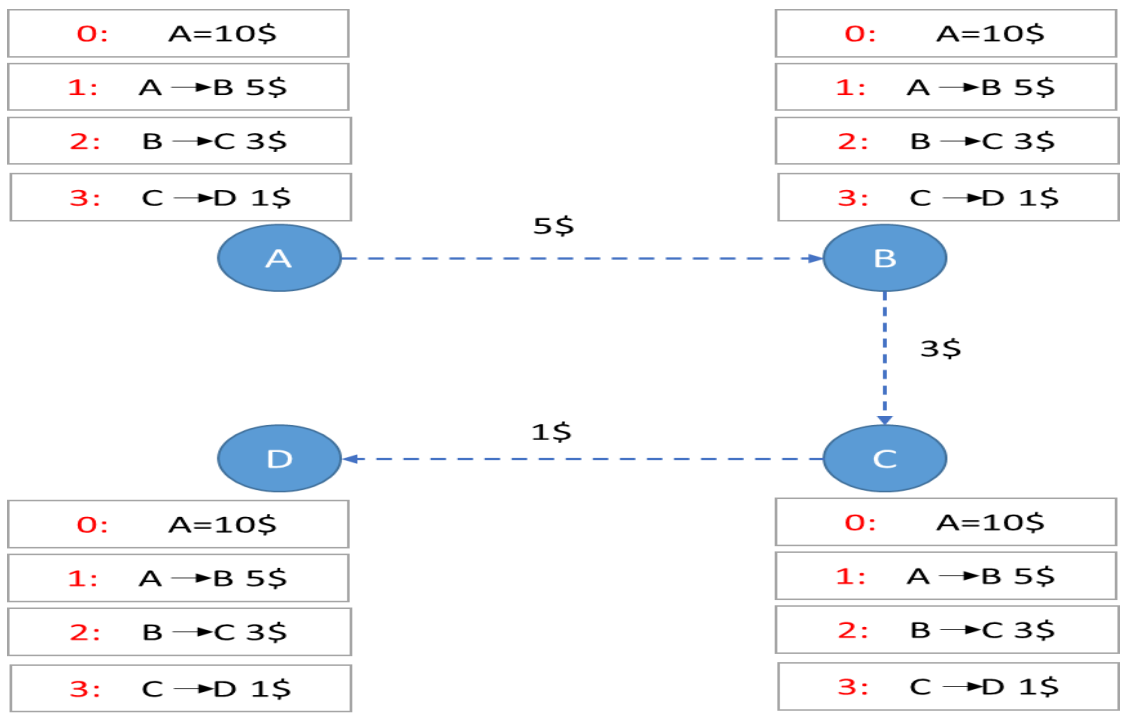

**Fig. 3**: The concept of the distributed ledger.

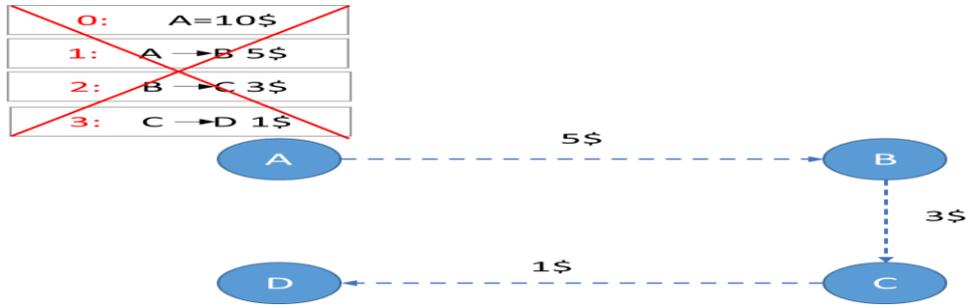

Fig. 4: Centralized place that holds the ledger.

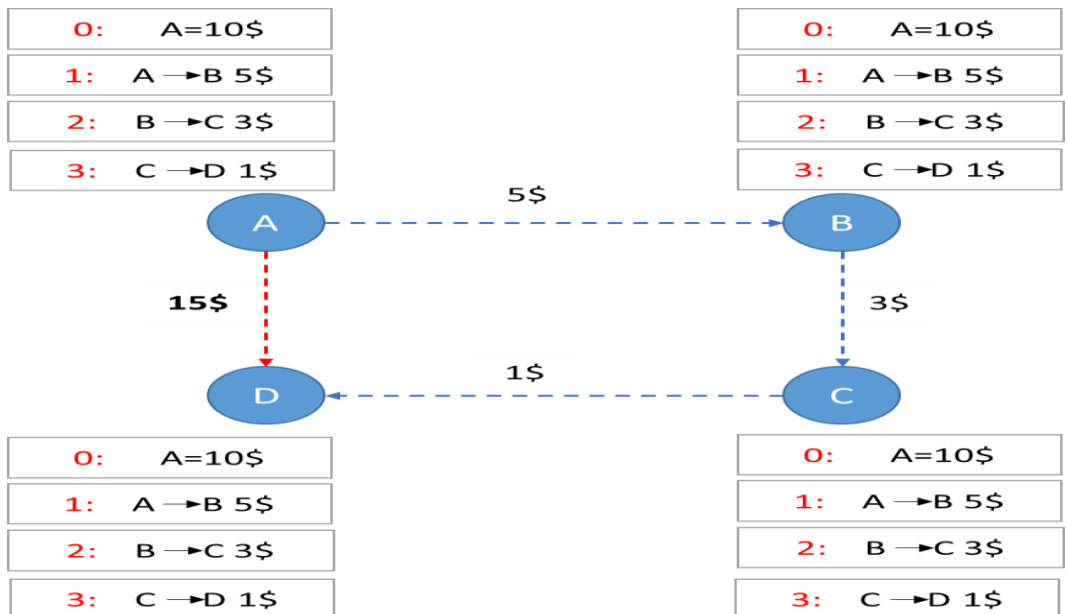

Fig. 5: Invalidated transaction

3. **Mining Process:** The network or the system should take into account the issue of synchronization, which all nodes must have the same copy of the ledger. This can be achieved by the third principle of BC which is the *Mining Process*. The following example explains how does mining process work.
**Example:** Suppose that node B wants to send five dollars to node D as shown in Fig. 6. Based on the open ledger previously mentioned, every node on the network will see that node B wants to send Five dollars to (B → D $5) as shown in Fig. 6, this is an invalidated transaction. The question is how does the system decide that this transaction is invalid. To answer this question, we need to explain the function of miners. Miners are special nodes in the system, which responsible for validating transactions and performing the consensus mechanism. In this case, suppose that nodes A and C are miners. Both nodes

A and C are going to perform the following two functions: The nodes A and B will compete among themselves who will be the first to take this transaction (B → C $5) and be able to validate it, which means that nod B has five dollars to send or no. After finishing the validation process, the miners A and B need to find a *special number*, which called a *nonce*. To find nonce, the miners need to invest significant computational power and time to search for the nonce.

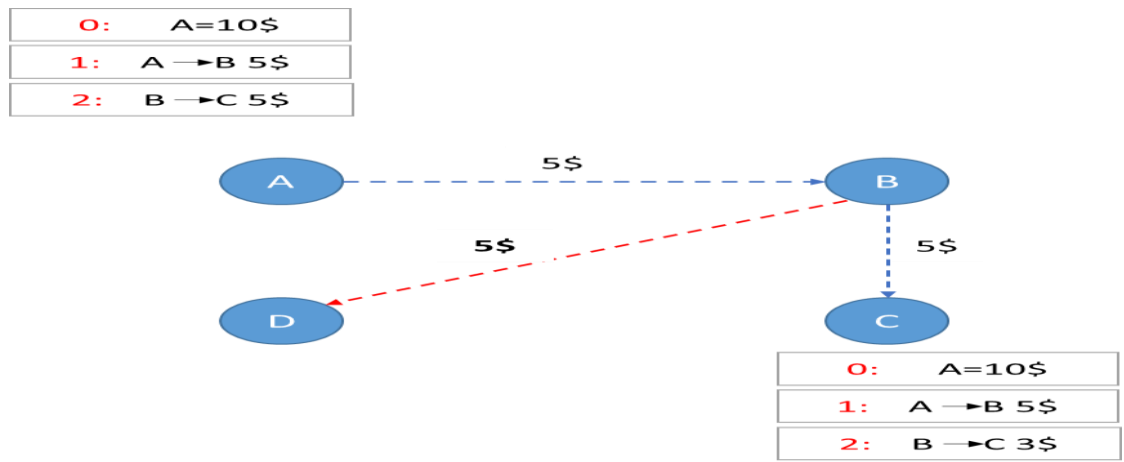

**Fig. 6**: Miners and mining process

## 2.2 What Does BlockChain Mean?

A BlockChain technology (BC) can be viewed as a decentralized, distributed, immutable, and shared a digital ledger that maintains a continuously growing list of blocks that are linked and secured using cryptography. From this definition the following four essential features of BC can be obtained:

1. **Decentralized:** BC is a decentralized system which means that there is no single point of control responsible for security, the control is shared through many independent entities such as computers, or enterprises, and managed by using private and public keys for encryption and decryption, digital digest, and hashing. To keep the decentralization going every BC must have a consensus algorithm to help the network make decisions or else the core value of it is lost.
2. **Distributed**: A decentralized system is also a distributed system so, the digital ledger is distributed and shared across many entities such as computers or organizations.
3. **Immutability**: Immutability is one of the most important features of BC technology which means that the blocks which contain data or transactions can't be modified retroactively without the modification of all subsequent blocks and the consensus of the network. Every participant on the system has a copy of the ledger and to add a transaction every participant needs to validate it. If the majority decides that it's valid, then its added to the digital ledger.

## 3. BlockChain Basics

### 3.1  BlockChain Architecture

BlockChain is a list of blocks, which contains transactions like traditional public ledger [17][18] as shown in Fig. 7. BlockChain consists of two types of elements:
1. Transactions: Transactions are the actions created by the participants in the network. The recorded transaction might be any information like payment history, personal data etc.
2. Blocks: Blocks record the transactions and make sure they are in the correct order and have not been tampered with. In the Bitcoin world, the block size is limited at 1 MB bytes, and the Average Transactions per block = Maximum Block Size / Average Transaction Size.

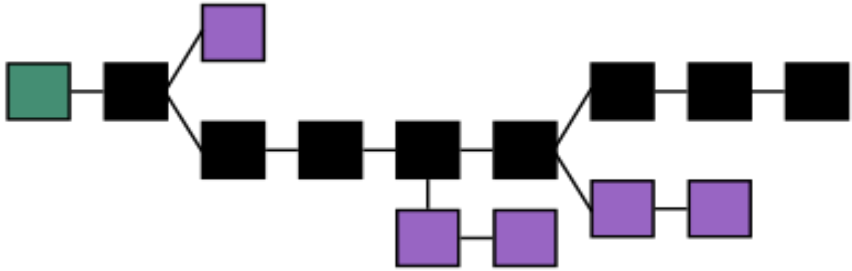

**Fig. 7:** BlockChain formation. Green: genesis block. Black: the main chain consists of the longest series of blocks from genesis to the current block. Purple: exist outside of the main chain [18].

Each block consists of block header and block data which is also called the transaction list as shown in Fig. 8. The block header contains the block number, a nonce value which is used by miners to solve the hash puzzle, hash of previous block header, hash of the current block and timestamp. Every block in the BC system is linked to the previous block using the hash of previous block header expect the first (genesis) block.

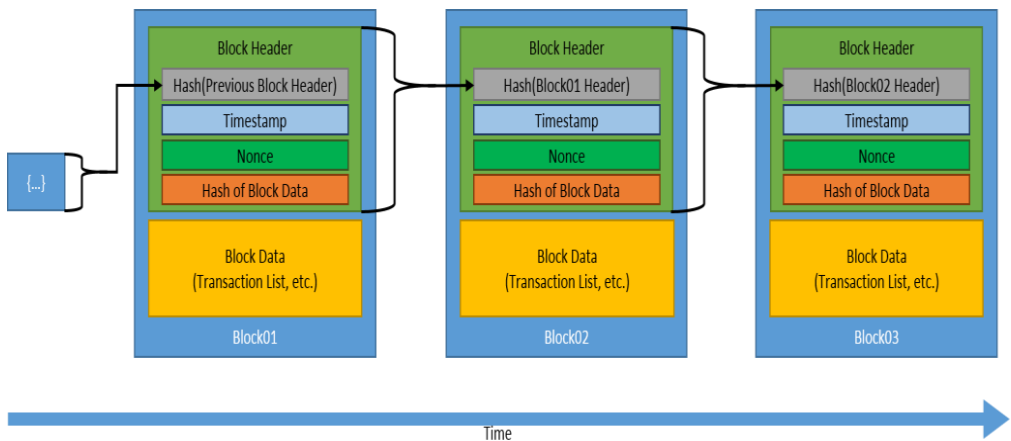

**Fig. 8**: BlockChain Structure [7].

The genesis block is the first block in any BC system. The second block to be added on top of genesis would then be referred to as block number 01. The number used to refer to the ordering of blocks is known as the block height number.

**3.2 Classification of BC**

Based on the data management, the can be classified into three main classes: Public, Private, and Consortium [19][20][21]. The public BC (Permissionless) is completely open and anyone can join and participate in the network. Some public BC limit access to just reading or writing. Bitcoin is one of the public BlockChain examples that permission to join it uses an approach where anyone can write. The second class is Private BC (Permission), which is not open to anyone and the participant needs permission to join it, because there is a restriction on who is allowed to participate in the network. This type of BC is used between companies that belong to the same legal mother entity [22]. Finally, the Consortium BC (Hybrid), which is would be a mix of both the public and private BC. Wherein the ability to read and write could be extended to a certain number of people or nodes. This could be used by groups of organizations or firms, who get together, work on developing different models by collaborating with each other.

**Table 2**: The differences between the three classes of BC

|  | **Private BC** | **Public BC** | **Hybrid BC** |
|---|---|---|---|
| Double Spending | Not applicable | Prohibited | Not applicable |
| Who is allowed to participate in the network? | The participant needs a permission to join the network. | Anyone can join the network. | The participant needs a permission to join the network. |
| Who is allowed to read, write and audit . | Preselected participant. | Any participant in the network. | Preselected participants |
| Miners | Preselected participant or participants. | Any participant can be a miner. | Preselected participant. |
| Finality | Supported | Not supported | Supported |
| Power Consumption | Low energy consumption | Large energy consumption | Low energy consumption |
| Scalability | A low to medium concern | A high concern | A low to medium concern |
| Decentralized | Centralized–some form of decentralization. | Fully decentralized | Centralized–some form of decentralization |
| Privacy | High level of privacy | No privacy | High level of privacy |
| Speed | Faster | Slower | Faster |
| Openness | Closed or open to a certain number preselected nodes. | Completely open. | Open to a certain number of preselected nodes. |

Table.2 shows the differences between the three classes of BC. The above table can help us to answer many frequently asked questions such as which BC should we use, as well as correct several misconceptions that most researchers or interested have about BC and distributed ledger technology.

### 3.3 How does BC Work?

It's always better to explain full scenario to understand more fully how does BC work, this scenario includes the following processes: Creation the transaction, Generation the keys, Performing the Singing Process (hashing & encryption), Broadcasting the transaction, Validation, and Mining Process (Performing a cryptographic puzzle/nonce). Once the transaction is created by any node or participant in the network, the process of generation private and public keys is started. Then the participant performs a so-called signing process which includes two steps: the first one is hashing the transaction using Sha256 which results a *hash value* and the resulted hash value encrypted using the private key to produce a digital signature. Once the digital signature is created, the participant starts broadcasting the transaction, digital signature and public key to all nodes (miners) in the network. As the Bitcoin is built on P2P, the broadcasting process is as follows: The participant first sends the *original transaction, digital signature and public key* to neighboring nodes, then those nodes do the same with their neighbors until the transaction reaches all nodes in the network.

Once the miners receive the transactions, they compete among themselves who will be the first to take this/these invalidated transaction(s) and be able to validate it/them and add it/them into a blockchain. The first miner that will do that will get a financial reward and in this case Bitcoin. In order to do that a miner needs to do two things:
- First needs to *validate* the transaction by performing the verification process to achieve integrity, authentication and non-repudiation.
- The second thing that a miner needs to do is to find a nonce (Applying a consensus algorithm such as Proof of work in the case of Bitcoin), the miners compete each other to solve a cryptographic puzzle. To find the nonce, the miner needs to invest computational power and time because the search for it is random. The miner is repeatedly guessing the nonce until it finds the key that match a random puzzle and publish a solution to the entire the network for adding it into the ledger.

### 3.4 Applications and Technical Challenges of BC Technology

The BC technology has the ability can be a panacea for many challenging problems across many businesses and industries, due to its features previously discussed in section 2. Fig. 9 shows some of the domains where BC finds applications and currently being used.

- **IoT Applications**: BC technology can be used in many domains, where IoT applications are involved, like sensing [23] [24], data storage [25] [26], identity management systems [27], healthcare applications [28], intelligent transportation systems [22], supply chain management [29] [30], and cyber law [31][32].

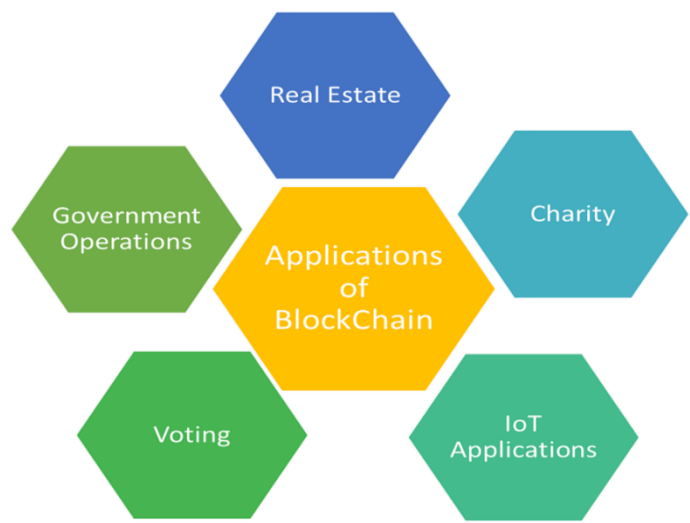

**Fig. 9**: Applications of BC

- **Voting**: For voting purposes the issue of transparency and integrity is considered as an important factor in determining whether it is valid or not. Thus, due to the fact that all the transactions are stored in a decentralized, distributed, and immutable system.
- **Government Operations**: Politics and transparency don't go really well with each other. The BC technology can increase the transparency and integrity by providing a decentralized digital ledger that can't be altered.
- **Real Estate:** The are many challenges regarding counterfeit during buying and selling real estate. Therefore, BC technology can reduce such events by making data reliable and faraway of being modified or lost.
- **Charity**: BC technology can address the issue of corruption in charitable organizations by allowing people to track their donations that they end up in the right hands.

### 3.5  BlockChain Implementation Challenges

In spite of the features of BC, it has some drawbacks as shown in Fig. 10

1. **Privacy:** BlockChain and privacy are two things that don't go together, which means that the full privacy is not the first concern for public BC. Private and hybrid BC can work with this level of privacy.

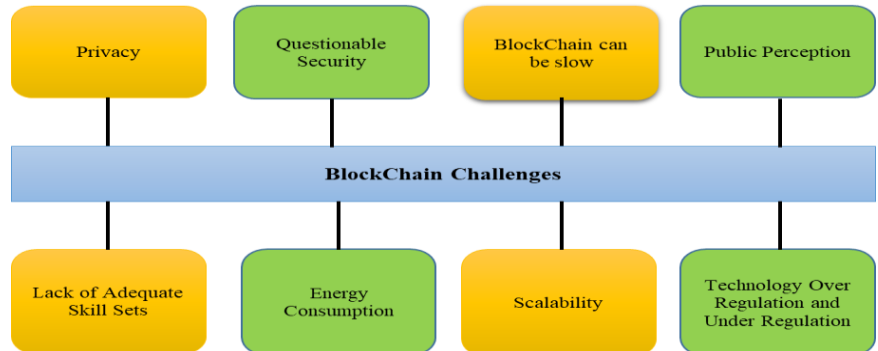

**Fig. 10**: Challenges of BC Technology.

2. **Questionable Security (public BC):** BC can address the security issue in terms of the data can't be modified but there is a case happened, when someone successfully gained control of 51% of the BC network and rolled back transactions to their own profit.
3. **Public BC can be slow:** Public BC needs long time to perform many operations such as singing, verification, encryption, decryption, and generation private and public keys.
4. **Lack of Qualified Candidates**: BC is a new technology and the employers claim that there are not enough qualified candidates in this field. Therefore, it is important to bridge the skills gap.
5. **Energy Consumption**: Public BC suffers from the energy consumption problem, because it consumes a high amount of energy to solve complex operations (mining process).
6. **Scalability**: BC system works excellent for a small-to-medium number of participants. But when the system is scaled up to thousands of users, the speed becomes a big challenge because the transitions take longer to process.
7. **Technology Over Regulations**: Till now there are not enough specifications about BC technology, which leads to break innovation and creativity in this field.

### 3.6 BlockChain for IoT Application

As previously mentioned in section 3.1 BlockChain technology can be used in many areas, where the IoT applications are involved [33]. However, it is important to note that a BC technology is not always the better solution for every IoT scenario. Therefore, in order to decide if the use of BC is appropriate for the certain IoT application or not, a designer or developer should take into consideration the following complementary aspects of IoT and BlockChain: Resources, Time consuming, Scalability, Latency, and Throughput. Mostly, IoT devices are resource-constrained with limited battery power, memory size, communication bandwidth, and processing performance, while BC devices are resource-consumed as shown in Table 3.

Table 3. Complementary aspects of IoT and BlockChain.

|  | BlockChain | IoT |
|---|---|---|
| **Resource** | Resource consuming. | Resource restricted. |
| **Scalability** | BC scale poorly with large networks. | IoT is expected to contain a large number of devices. |
| **Bandwidth** | BC has high bandwidth consumption. | IoT devices have limited bandwidth. |
| **Latency** | There is a significant delay associated with ensuring that a transaction is confirmed by the network. | Most IoT devices have stricter delay requirements. |
| **Throughput** | The throughput is defined as the number of transactions that can be stored. BC have limited throughput about 10 transaction per second | The number of transactions in the IoT would far exceed such limits due to extensive interactions between various platforms. |

BlockChain technology offers a way of recording transactions in a way that is designed to be secure, transparent, highly resistant to outages, auditable, and efficient [16] Consequently, a BC can be one of the solutions for addressing the security and privacy issues by providing:
1. A distributed and immutable system across a network.
2. Hash-based security, verification of identity and provenance authentication.
3. Consensus and agreement algorithms such as Proof of Work (POW) or Proof of Stake (POS) [33] for detecting bad users and mitigating threats.

The BlockChain, on the other hand can benefit from the IoT by extending its scope to deal with other real applications in more distributed and dynamic manner. In addition to that some IoT applications may require to perform economic transactions with third parties, therefore, BC can be used to address this issue.

## 4. Conclusion

BlockChain is a new technology that came without date. It might not be very attractive right now and any new technology will always go through a hype stage. It takes a lot of time to get rid of all the challenges and use it to power the modern world. Therefore, many researchers have suggested the integration of IoT with BC technology to address security challenges. This paper dealt the role of BC technology to support several IoT applications such as voting, health care, real estate, charity, etc. to overcome the lacks of IoT devices regarding security issues due to the fact that BC technology is an immutable system across a network, uses hash function and consensus algorithms for detecting bad users and mitigating threats. I hope this paper can serve as a primer for researchers to understand the concepts of BC technology, how BC relates to Bitcoin, what kinds of problems it solves and the importance of integration of BC with IoT for addressing the IoT security issues in order to develop BC-based IoT applications.